\documentclass[letterpaper,ngerman,english,aps,showpacs,showkeys]{revtex4-1}
\usepackage[T1]{fontenc}
\usepackage[latin9]{inputenc}
\usepackage{amsmath}
\usepackage{amssymb}



\usepackage{babel}
\begin{document}

\title{Two-dimensional Klein-Gordon Oscillator in the presence of a minimal
length }

\author{Abdelmalek Boumali}
\email{boumali.abdelmalek@gmail.com}

\affiliation{Laboratoire de Physique Appliquée et Théorique, ~\\
 Université Larbi-Tébessi- Tébessa, Algeria.}

\author{Zina Selama}
\email{zinaslama@gmail.com}

\affiliation{Laboratoire de Physique Appliquée et Théorique, ~\\
 Université Larbi-Tébessi- Tébessa, Algeria.}
\begin{abstract}
Minimal length of a two-dimensional Klein-Gordon oscillator is investigated
and illustrates the wave functions in the momentum space. The energy
eigenvalues are found and the corresponding wave functions are calculated
in terms of hyper-geometric functions.  
\end{abstract}
\maketitle

\section{introduction}

Recently, there have been growing interest in obtaining exact solutions
of relativistic wave equations. In particular exact solutions of the
Klein\textendash Gordon equation with various vector and scalar potentials. 

The Dirac relativistic oscillator is an important potential both for
theory and application. It was for the first time studied by Ito et
al\citep{1}. They considered a Dirac equation in which the momentum
$\vec{p}$ is replaced by $\vec{p}-im\beta\omega\vec{r}$, with $\vec{r}$
being the position vector, $m$ the mass of particle, and $\omega$
the frequency of the oscillator. The interest in the problem was revived
by Moshinsky and Szczepaniak \citep{2}, who gave it the name of Dirac
oscillator (DO) because, in the non-relativistic limit, it becomes
a harmonic oscillator with a very strong spin-orbit coupling term.
Physically, it can be shown that the (DO) interaction is a physical
system, which can be interpreted as the interaction of the anomalous
magnetic moment with a linear electric field \citep{3,4}. The electromagnetic
potential associated with the DO has been found by Benitez et al \citep{5}.
The Dirac oscillator has attracted a lot of interest both because
it provides one of the examples of the Dirac's equation exact solvability
and because of its numerous physical applications (see \citep{6}
and reference therein). Recently, Franco-Villafane et al \citep{7}
exposed the proposal of the first experimental microwave realization
of the one-dimensional DO. The experiment relies on a relation of
the DO to a corresponding tight-binding system. The experimental results
obtained, concerning the spectrum of the one-dimensional DO with and
without the mass term, are in good agreement with those obtained in
the theory. In addition, Quimbay et al \citep{8,9} show that the
Dirac oscillator can describe a naturally occurring physical system.
Specifically, the case of a two-dimensional Dirac oscillator can be
used to describe the dynamics of the charge carriers in graphene,
and hence its electronic properties. Also, the exact mapping of the
DO in the presence of a magnetic field with a quantum optics leads
to consider the DO as a theory of an open quantum systems coupled
to a thermal bath ( see Ref \citep{6} and references therein).

The unification between the general theory of relativity and the quantum
mechanics is one of the most important problems in theoretical physics.
This unification predicts the existence of a minimal measurable length
on the order of the Planck length. All approaches of quantum gravity
show the idea that near the Planck scale, the standard Heisenberg
uncertainty principle should be reformulated. The minimal length uncertainty
relation has appeared in the context of the string theory, where it
is a consequence of the fact that the string cannot probe distances
smaller than the string scale $\hbar\sqrt{\beta}$, where $\beta$
is a small positive parameter called the deformation parameter. This
minimal length can be introduced as an additional uncertainty in position
measurement, so that the usual canonical commutation relation between
position and momentum operators becomes$\left[\hat{x},\hat{p}\right]=i\hbar\left(1+\beta p^{2}\right)$.
This commutation relation leads to the standard Heisenberg uncertainty
relation $\triangle\hat{x}\triangle\hat{p}\geq i\hbar\left(1+\beta\left(\triangle p\right)^{2}\right)$,
which clearly implies the existence of a non-zero minimal length $\triangle x_{\mbox{min}}=\hbar\sqrt{\beta}$.
This modification of the uncertainty relation is usually termed the
generalized uncertainty principle (GUP) or the minimal length uncertainty
principle\citep{11,12,13,14}.

Nowadays, the reconsideration of the relativistic quantum mechanics
in the presence of a minimal measurable length have been studied extensively.
In this context, many papers were published where a different quantum
system in space with Heisenberg algebra was studied. They are: the
Abelian Higgs model \citep{15}, the thermostatics with minimal length
\citep{16}, the one-dimensional Hydrogen atom \citep{17}, the casimir
effect in minimal length theories \citep{18}, the effect of minimal
lengths on electron magnetism \citep{19}, the Dirac oscillator in
one and three dimensions \citep{20,21}, the solutions of a two-dimensional
Dirac equation in presence of an external magnetic field \citep{22},
the non-commutative phase space Schrodinger equation\citep{23}, Schrodinger
equation with Harmonic potential in the presence of a Magnetic Field
\citep{24}.

The purpose of this work is to investigate the formulation of a two-dimensional
Klein-Gordon oscillator by solving fundamental equations in the framework
of relativistic quantum mechanics with minimal length. The problem
describes a relativistic particle moving in the relativistic harmonic
oscillator called the Dirac oscillator.

The paper is organized as follows. In section II, we exposed the solutions
of our problem within habitual quantum mechanics, using the new method
developed by Menculini et al \citep{10} and Jana et al\citep{25}.
Then, section III will be devoted to the our case, i.e., the solution
of a two dimensional Klein-Gordon oscillator in the framework of relativistic
quantum mechanics with minimal length. Finally, in section V, we present
the conclusion.

\section{The solutions within habitual quantum mechanics}

A two-dimensional Klein-Gordon oscillator is 
\begin{equation}
\left\{ \underbrace{\left(p_{x}+im_{0}\omega x\right)\left(p_{x}-im_{0}\omega x\right)}+\underset{\cap}{\underbrace{\left(p_{y}+im_{0}\omega y\right)\left(p_{y}-im_{0}\omega y\right)}}-\frac{E^{2}-m_{0}^{2}c^{4}}{c^{2}}\right\} \psi_{KG}=0,\label{eq:1}
\end{equation}
with
\begin{align}
\cup & =\left(p_{x}+im_{0}\omega x\right)\left(p_{x}-im_{0}\omega x\right)\nonumber \\
 & =p_{x}^{2}+m_{0}^{2}\omega^{2}x^{2}-m_{0}\omega\hbar,\label{eq:2}
\end{align}
\begin{align}
\cap & =\left(p_{y}+im_{0}\omega y\right)\left(p_{y}-im_{0}\omega y\right)\nonumber \\
 & =p_{x}^{2}+m_{0}^{2}\omega^{2}x^{2}-m_{0}\omega\hbar.\label{eq:3}
\end{align}
Now, for the sake of simplicity, we bring the problem into the momentum
space. Recalling that 
\begin{equation}
x=i\hbar\frac{\partial}{\partial p_{x}},\,y=i\hbar\frac{\partial}{\partial p_{x}},\label{eq:4}
\end{equation}
\begin{equation}
\hat{p}_{x}=p_{x},\,\hat{p}_{y}=p_{y},\label{eq:5}
\end{equation}
and passing onto polar coordinates with the following definition \citep{10}
\begin{equation}
p_{x}=p\cos\theta,\,p_{y}=p\sin\theta,\,\text{with}\,p^{2}=p_{x}^{2}+p_{y}^{2},\label{eq:6}
\end{equation}
\begin{equation}
\hat{x}=i\hbar\frac{\partial}{\partial p_{x}}=i\hbar\left(\cos\theta\frac{d}{dp}-\frac{\sin\theta}{p}\frac{d}{d\theta}\right),\label{eq:7}
\end{equation}
\begin{equation}
\hat{y}=i\hbar\frac{\partial}{\partial p_{y}}=i\hbar\left(\sin\theta\frac{d}{dp}+\frac{\cos\theta}{p}\frac{d}{d\theta}\right),\label{eq:8}
\end{equation}
Eqs. (\ref{eq:2}) and (\ref{eq:3}) transform into
\begin{equation}
\cup=p^{2}\cos^{2}\theta-m_{0}^{2}\omega^{2}\hbar^{2}\frac{\partial^{2}}{\partial p_{x}^{2}}-\hbar m_{0}\omega,\label{eq:9}
\end{equation}
\begin{equation}
\cap=p^{2}\sin^{2}\theta-m_{0}^{2}\omega^{2}\hbar^{2}\frac{\partial^{2}}{\partial p_{y}^{2}}-\hbar m_{0}\omega.\label{eq:10}
\end{equation}
In this case, Eq. (\ref{eq:1}) becomes
\begin{equation}
\left(p^{2}-m_{0}^{2}\omega^{2}\hbar^{2}\left(\frac{\partial^{2}}{\partial p_{x}^{2}}+\frac{\partial^{2}}{\partial p_{y}^{2}}\right)-2\hbar m_{0}\omega-\varsigma\right)\psi_{KG}=0.\label{eq:11}
\end{equation}
By using Eqs. (\ref{eq:6}), (\ref{eq:7}) and (\ref{eq:8}), we have
\begin{equation}
\frac{\partial^{2}}{\partial p_{x}^{2}}+\frac{\partial^{2}}{\partial p_{y}^{2}}=\frac{\partial^{2}}{\partial p^{2}}+\frac{1}{p^{2}}\frac{\partial^{2}}{\partial\vartheta^{2}}+\frac{1}{p}\frac{\partial}{\partial p},\label{eq:12}
\end{equation}
and consequently, we obtain
\begin{equation}
\left\{ p^{2}-\lambda^{2}\left(\frac{\partial^{2}}{\partial p^{2}}+\frac{1}{p^{2}}\frac{\partial^{2}}{\partial\theta^{2}}+\frac{1}{p}\frac{\partial}{\partial p}\right)-2\lambda-\varsigma\right\} \psi_{KG}=0,\label{eq:13}
\end{equation}
with $\lambda=m\omega\hbar$ and
\begin{equation}
\varsigma=\frac{E^{2}-m_{0}^{2}c^{4}}{c^{2}}.\label{eq:14}
\end{equation}
Here we have used that
\begin{align*}
\frac{\partial}{\partial p_{x}} & =\frac{\partial}{\partial p}\frac{\partial p}{\partial p_{x}}+\frac{\partial}{\partial\theta}\frac{\partial\theta}{\partial p_{x}}==\cos\theta\frac{\partial}{\partial p}-\frac{\sin\theta}{p}\frac{\partial}{\partial\theta},
\end{align*}
\begin{align}
\frac{\partial}{\partial p_{y}} & =\frac{\partial}{\partial p}\frac{\partial p}{\partial p_{y}}+\frac{\partial}{\partial\theta}\frac{\partial\theta}{\partial p_{y}}=\text{sin}\theta\frac{\partial}{\partial p}+\frac{\text{cos}\theta}{p}\frac{\partial}{\partial\theta}.\label{eq:14-2}
\end{align}
Now, when we choose 
\begin{equation}
\psi_{KG}\left(p,\theta\right)=f\left(p\right)e^{i\left|l\right|\theta},\label{eq:15}
\end{equation}
Eq. (\ref{eq:13}) is transformed into 
\begin{equation}
\left(\frac{d^{2}f\left(p\right)}{dp^{2}}+\frac{1}{p}\frac{df\left(p\right)}{dp}-\frac{l^{2}}{p^{2}}f\left(p\right)\right)+\left(\kappa^{2}-k^{2}p^{2}\right)f\left(p\right)=0,\label{eq:16}
\end{equation}
with 
\begin{equation}
\kappa^{2}=\frac{2\lambda+\zeta}{\lambda^{2}},\,k^{2}=\frac{1}{\lambda^{2}}.\label{eq:17}
\end{equation}
Now, Putting\citep{25} 
\begin{equation}
f\left(p\right)=p^{\left|l\right|}e^{-\frac{k}{2}p^{2}}F\left(p\right),\label{eq:18}
\end{equation}
the differential equation 
\begin{equation}
F^{''}+\left(\frac{2\left|l\right|+1}{p}-2kp\right)F^{'}-\left[2k\left(\left|l\right|+1\right)-\kappa^{2}\right]F=0,\label{eq:19}
\end{equation}
is obtained for $F\left(p\right)$ which by using, instead of $p$,
the variable $xt=kp^{2}$, is transformed into the Kummer equation
\begin{equation}
t\frac{d^{2}F}{dt^{2}}+\left\{ \left|l\right|+1-t\right\} \frac{dF}{dt}-\frac{1}{2}\left\{ \left|l\right|+1-\frac{\kappa^{2}}{4k}\right\} F=0,\label{eq:20}
\end{equation}
whose solution is the confluent series $_{1}F_{1}\left(a;\left|l\right|+1;t\right)$,
with 
\begin{equation}
a=\frac{1}{2}\left(\left|l\right|+1\right)-\frac{\kappa^{2}}{4k}.\label{eq:21}
\end{equation}
The confluent series becomes a polynomial if and only if $a=-n,\,\left(n=0,1,2,\right)$.
We then have the solutions 
\begin{equation}
\psi_{KG}\left(p,\theta\right)=C_{n,\left|l\right|}p^{\left|l\right|}e^{-\frac{k}{2}p^{2}}\,_{1}F_{1}\left(-n;\left|l\right|+1;kp^{2}\right)e^{i\left|l\right|\theta},\label{eq:22}
\end{equation}
\begin{equation}
E_{n}=\pm mc_{0}^{2}\sqrt{1+2rN}.\label{eq:23}
\end{equation}
with $N=2n+\left|l\right|$ is the principal quantum number, and $r=\frac{\hbar\omega}{m_{0}c^{2}}$.
This form of energy is in a good agreement with that obtained in the
literature \citep{11}. 

\section{The solutions in the presence of a minimal length}

In the minimal length formalism, the Heisenberg algebra is given by
\citep{12,13,14,15,16,17,18,19,20,21,22,23,24}
\begin{equation}
\left[\hat{x}_{i},\hat{p}_{i}\right]=i\hbar\delta_{ij}\left(1+\beta p^{2}\right),\label{eq:38}
\end{equation}
where $\beta>0$ is the minimal length parameter. A representation
of $\hat{x}_{i}$ and $\hat{p}_{i}$ which satisfies Eq. (\ref{eq:38}),
may be taken as 
\begin{equation}
\hat{x}=i\hbar\left(1+\beta p^{2}\right)\frac{d}{dp_{x}},\,\hat{p}_{x}=p_{x},\label{eq:39}
\end{equation}
\begin{equation}
\hat{y}=i\hbar\left(1+\beta p^{2}\right)\frac{d}{dp_{y}},\,\hat{p}_{y}=p_{y}.\label{eq:40}
\end{equation}
In this case, the KG oscillator equation is
\begin{equation}
\left\{ p_{x}^{2}+p_{y}^{2}+m^{2}\omega^{2}\left(x^{2}+y^{2}\right)+im\omega\left[x,p_{x}\right]+im\omega\left[y,p_{y}\right]-\varsigma\right\} \psi_{KG}=0.\label{eq:41}
\end{equation}
By using the equations (\ref{eq:39}) and (\ref{eq:40}), Eq. (\ref{eq:41})
becomes
\begin{equation}
\left\{ p^{2}-\lambda^{2}\left(1+\beta p^{2}\right)^{2}\left(\frac{\partial^{2}}{\partial p^{2}}+\frac{1}{p^{2}}\frac{\partial^{2}}{\partial\theta^{2}}+\frac{1}{p}\frac{\partial}{\partial p}\right)-2\beta\lambda^{2}\left(1+\beta p^{2}\right)^{2}p\frac{\partial}{\partial p}-2\lambda\left(1+\beta p^{2}\right)-\varsigma\right\} \psi_{KG}=0.\label{eq:42}
\end{equation}
\foreignlanguage{ngerman}{Now, when we put that
\begin{equation}
\psi_{KG}=h(p)e^{i\left|j\right|\theta},\label{eq:43}
\end{equation}
with $j=0,\pm1,\pm2,\ldots$, then Eq. (\ref{eq:42}) transformed
into
\begin{equation}
\left\{ -a(p)\frac{\partial^{2}}{\partial p^{2}}+b(p)\frac{\partial}{\partial p}+c(p)-\varsigma\right\} h(p)=0,\label{eq:44}
\end{equation}
with
\begin{equation}
\begin{array}{c}
a(p)=\lambda^{2}\left(1+\beta p^{2}\right)^{2},\\
b(p)=-\frac{\lambda^{2}\left(1+\beta p^{2}\right)^{2}}{p}-2\beta\lambda^{2}\left(1+\beta p^{2}\right)p=-\frac{a}{p}-2\beta\lambda\sqrt{a}p\\
c(p)=p^{2}+\frac{j^{2}\lambda^{2}\left(1+\beta p^{2}\right)^{2}}{p^{2}}-2\lambda\left(1+\beta p^{2}\right)=p^{2}+j^{2}\frac{a}{p^{2}}-2\sqrt{a}.
\end{array},\label{eq:45}
\end{equation}
Now, according to the following substitution \citep{25}
\begin{equation}
h(p)=\rho(p)\varphi(p),\;q=\int\frac{1}{\sqrt{a(p)}}dp,\label{eq:46}
\end{equation}
\begin{equation}
\rho(p)=\exp\left(\int\chi(p)dp\right),\,\chi\left(p\right)=\frac{2b+a'}{4a}=-\frac{1}{2p},\label{eq:47}
\end{equation}
we have
\begin{equation}
\left[-\frac{d^{2}\varphi(p)}{dq^{2}}+V(p)\right]\varphi(p)=\varsigma\varphi(p),\label{eq:48}
\end{equation}
with
\begin{equation}
V(p)=p^{2}-2\lambda\left(1+\beta p^{2}\right)+\beta\lambda^{2}\left(1+\beta p^{2}\right)+\frac{\lambda^{2}\left(1+\beta p^{2}\right)^{2}}{p^{2}}\left(j^{2}-\frac{1}{4}\right).\label{eq:49}
\end{equation}
We note here that the function $\rho(p)=p^{-\frac{1}{2}}$.}

\selectlanguage{ngerman}%
Now, if we put
\begin{equation}
p=\frac{1}{\sqrt{\beta}}\tan\left(q\lambda\sqrt{\beta}\right),\label{eq:50}
\end{equation}
the term $V\left(p\right)$ becomes
\begin{equation}
V\left(p\right)=-\frac{1}{\beta}+\underbrace{\beta\lambda^{2}}_{U_{0}}\left(\frac{j^{2}+\frac{3}{4}-\frac{1}{\beta\lambda}+\frac{1}{\beta^{2}\lambda^{2}}}{\cos^{2}\alpha q}+\frac{j^{2}-\frac{1}{4}}{\sin^{2}\alpha q}\right),\label{eq:51}
\end{equation}
and consequently, the final form of our differential equation is
\begin{equation}
\left\{ -\frac{d^{2}\varphi(p)}{dq^{2}}+\frac{U_{0}}{2}\left(\frac{j^{2}+\frac{3}{4}-\frac{2}{\beta\lambda}+\frac{1}{\beta^{2}\lambda^{2}}}{\cos^{2}\alpha q}+\frac{j^{2}-\frac{1}{4}}{\sin^{2}\alpha q}\right)\right\} \varphi(p)=\bar{\varsigma}\varphi(p),\label{eq:52}
\end{equation}
where $\bar{\varsigma}=\varsigma+\frac{1}{\beta}$.\foreignlanguage{english}{{\small{}
Eq. (\ref{eq:52}) brings into 
\begin{equation}
\left\{ -\frac{d^{2}\varphi(p)}{dq^{2}}+\frac{U_{0}}{2}\left(\frac{\zeta_{1}\left(\zeta_{1}-1\right)}{\cos^{2}\alpha q}+\frac{\zeta_{2}\left(\zeta_{2}-1\right)}{\sin^{2}\alpha q}\right)\right\} \varphi(p)=\bar{\varsigma}\varphi(p)\label{eq:53}
\end{equation}
}where
\begin{equation}
V\left(p\right)=-\frac{1}{\beta}+\beta\lambda^{2}\left\{ \frac{\zeta_{1}\left(\zeta_{1}-1\right)}{\sin^{2}\left(\alpha q\right)}+\frac{\zeta_{2}\left(\zeta_{2}-1\right)}{\cos^{2}\left(\alpha q\right)}\right\} ,\label{eq:54}
\end{equation}
with 
\begin{equation}
\zeta_{1}\left(\zeta_{1}-1\right)=j^{2}-\frac{1}{4},\label{eq:55}
\end{equation}
\begin{equation}
\zeta_{2}\left(\zeta_{2}-1\right)=j^{2}+\frac{3}{4}-\frac{2}{\beta\lambda}+\frac{1}{\beta^{2}\lambda^{2}}.\label{eq:56}
\end{equation}
Thus, we have 
\begin{equation}
\left(-\frac{d^{2}}{dq^{2}}+\frac{1}{2}U_{0}\left\{ \frac{\zeta_{1}\left(\zeta_{1}-1\right)}{\sin^{2}\left(\alpha q\right)}+\frac{\zeta_{2}\left(\zeta_{2}-1\right)}{\cos^{2}\left(\alpha q\right)}\right\} \right)\varphi\left(p\right)=\bar{\xi}^{2}\varphi\left(p\right),\label{eq:57}
\end{equation}
where $U_{0}=\alpha^{2}$ with $\alpha=\lambda\sqrt{\beta}$. Eq.
(\ref{eq:57}) is the well-known Schrodinger equation in a P\"{o}schl-Teller
potential with \citep{25} 
\begin{equation}
U=\frac{1}{2}U_{0}\left\{ \frac{\zeta_{1}\left(\zeta_{1}-1\right)}{\sin^{2}\left(\alpha q\right)}+\frac{\zeta_{2}\left(\zeta_{2}-1\right)}{\cos^{2}\left(\alpha q\right)}\right\} ,\label{eq:58}
\end{equation}
and with the following conditions $\zeta_{1}>1$ and $\zeta_{2}>1$.}

\selectlanguage{english}%
By comparison Eq. (\ref{eq:52}) with Eq. (\ref{eq:57}), we have
\begin{equation}
\zeta_{1}=\left|j\right|\pm\frac{1}{2},\label{eq:59}
\end{equation}
\begin{equation}
\zeta_{2}=\frac{1}{2}\pm\left(\frac{1}{\beta\lambda}-1\right)\sqrt{1+\frac{j^{2}}{\left(\frac{1}{\beta\lambda}-1\right)^{2}}}.\label{eq:60}
\end{equation}
Now, in order to solve Eq. (\ref{eq:52}), we introduce the new variable
\begin{equation}
z=\sin^{2}\left(\alpha q\right).\label{eq:61}
\end{equation}
In this case, Eq. (\ref{eq:52}) can be written by 
\begin{equation}
z\left(1-z\right)\varphi^{''}+\left(\frac{1}{2}-z\right)\varphi^{'}+\frac{1}{4}\left\{ \frac{\bar{\xi}^{2}}{\alpha^{2}}-\frac{\zeta_{1}\left(\zeta_{1}-1\right)}{z}-\frac{\zeta_{2}\left(\zeta_{2}-1\right)}{1-z}\right\} \varphi=0.\label{eq:62}
\end{equation}
With the new wave function $\varphi$, defined by 
\begin{equation}
\varphi=z^{\frac{\zeta_{1}}{2}}\left(1-z\right)^{\frac{\zeta_{2}}{2}}\Psi\left(z\right),\label{eq:63}
\end{equation}
we arrive at 
\begin{equation}
z\left(1-z\right)\Psi^{''}+\left[\left(\zeta_{1}+\frac{1}{2}\right)-z\left(\zeta_{1}+\zeta_{2}+1\right)\right]\Psi^{'}+\frac{1}{4}\left\{ \frac{\bar{\xi}^{2}}{\alpha^{2}}-\left(\zeta_{1}+\zeta_{2}\right)^{2}\right\} \Psi=0.\label{eq:64}
\end{equation}
The general solution of this equation is 
\begin{equation}
\Psi=C_{1}\,_{2}F_{1}\left(a;b;c;z\right)+C_{2}\,z^{1-c}\,_{2}F_{1}\left(a+1-c;b+1-c;2-c;z\right),\label{eq:65}
\end{equation}
with 
\begin{equation}
a=\frac{1}{2}\left(\zeta_{1}+\zeta_{2}+\frac{\bar{\xi}}{\alpha^{2}}\right),\,b=\frac{1}{2}\left(\zeta_{1}+\zeta_{2}-\frac{\bar{\xi}}{\alpha^{2}}\right),\,c=\zeta_{1}+\frac{1}{2}.\label{eq:66}
\end{equation}
With the condition $a=-n$, we obtain 
\begin{equation}
\bar{\xi}^{2}=\alpha^{2}\left(\zeta_{1}+\zeta_{2}+2n\right)^{2}.\label{eq:67}
\end{equation}
In order to obtain the energy spectrum, it should be to note that
in the limit $\beta\rightarrow0$, the energy spectrum should be covert
to no-GUP result.

Thus, the exact form of $\zeta_{1}$ and $\zeta_{2}$ are 
\begin{equation}
\zeta_{1}=\left|j\right|+\frac{1}{2},\label{eq:68}
\end{equation}
\begin{equation}
\zeta_{2}=\frac{1}{2}+\frac{1}{2}+\left(\frac{1}{\beta\lambda}-1\right)\sqrt{1+\frac{j^{2}}{\left(\frac{1}{\beta\lambda}-1\right)^{2}}}.\label{eq:69}
\end{equation}
where $j\neq0$.

With the aid of Eqs. (\ref{eq:59}), (\ref{eq:60}) and (\ref{eq:67}),
we obtain the final form of the spectrum of energy, which is given
by 
\begin{equation}
\frac{E}{m_{0}c^{2}}=\pm\sqrt{1-2r+2\Sigma r\left(N+1\right)+\frac{\beta}{\beta_{0}}r^{2}\left(N^{2}-2\Sigma\left(N+1\right)+j^{2}\right)},\label{eq:70}
\end{equation}
with
\begin{equation}
\Sigma=\sqrt{1+\frac{j^{2}}{\left(\frac{1}{\beta\lambda}-1\right)^{2}}}.\label{eq:70.1}
\end{equation}
We can see that where $\beta\rightarrow0,$ $\Sigma\rightarrow1$,
and the energy spectrum is covert to no-GUP result (Eq. (\ref{eq:23})).

The corresponding wave function is 
\begin{equation}
\psi_{KG}=Ne^{i\left|j\right|\theta}p^{-\frac{1}{2}}z^{\frac{\zeta_{1}}{2}}\left(1-z\right)^{\frac{\zeta_{2}}{2}}\,_{2}F_{1}\left(a;b;c;z\right).\label{eq:71}
\end{equation}
with $N$ is the constant of normalization.

\section{Conclusion}

In this paper, we have exactly solved the Klein-Gordon oscillator
in two dimensions in the framework of relativistic quantum mechanics
with minimal length. The eigensolutions are obtained using a method
developed in refs. \citep{10,25} to solve a two-dimensional Dirac
equation and Klein-Gordon equations. We firstly consider the case
of the Klein-Gordon oscillator within the ordinary quantum mechanics:
our results are in good agreement with those obtained in the literature.
Then we have extended it in the case of the presence of a minimal
length. The energy levels, for both cases, show a dependence on $N^{2}$
in the presence of the minimal length, which describes a hard confinement.
In the limit where $\beta\rightarrow0$, we recover the energy spectrum
of no-GUP.

\end{document}